# ASIA: An Access Control, Session Invocation and Authorization Architecture for Home Energy Appliances in Smart Energy Grid Environments

Rainer Falk, Steffen Fries, Hans Joachim Hof
Siemens AG
Corporate Technology
Germany
{rainer.falk; steffen.fries; hans-joachim.hof}@siemens.com

*Abstract*—**With the advent of the smart energy grid – an energy transportation and distribution network being combined with an IT network for its monitoring and control – information security has gained tremendous importance for energy distribution and energy automation systems. Integrated security functionality is crucial to ensure a reliable and continuous operation of the smart energy grid. Further security related challenges arise from the integration of millions of smart homes into the smart grid. This paper gives an overview of the smart energy grid environment and its challenges. Many future use cases are centered around the smart home, using an ICT gateway. Approaches to protect the access and data exchange are described, preventing manipulation of ICT gateway operation. The paper presents ASIA – an Authentication, Session Invocation, and Authorization component to be used in the smart energy grid, to protect ICT gateways and to cope with problems like ICT gateway discovery and ICT gateway addressing.**

## I. Introduction

Power generation, transmission, and distribution systems are characterized by the existence of two parallel infrastructures, the electrical grid carrying the energy, and the information and communication infrastructure used to automate, control and supervise the electrical grid. Especially the latter is becoming more and more one of the essential parts for power system operations as it is responsible not only for retrieving information from field equipment but most importantly for transmitting control commands. The power system as a critical infrastructure requires integrated protection against intentional attacks. A highly dependable management and operation of the information infrastructure is a prerequisite for a highly reliable energy network as the power system relies on the availability of the information infrastructure. Therefore the information infrastructure must be operated according to the same level of reliability as required for the stability of the power system infrastructure to prevent any type of outage. Integrated security functionality is therefore a central part of the energy control network.

Today's rather centralized approach for power generation has already and will continue to evolve into a decentralized power generation involving existing power plants, power plants producing renewable energy (like wind parks) down to households having their own micro power plants (e.g., solar cells). The importance of decentralized energy generation is expected to increase in the future due to the efforts to reduce the $CO_2$ footprint to fight global warming. Introducing decentralized energy generators into the current energy distribution network poses great challenges for energy automation (EA) in the smart grid scenario, especially secure communication between a control station (e.g., substation) and equipment of users (e.g., decentralized energy generators) must be addressed. Another trend in energy automation networks is the extended focus on the customer/user. From an architectural approach, users are expected to connect to the smart grid or more specific to the information and communication technology (ICT) infrastructure of the smart grid using an ICT gateway, which on the one hand provides metering data to the meter management service based on the locally measured and accumulated data. On the other hand, the ICT gateway also receives commands for load reductions to avoid power demand peaks. Additionally, an interface for the user to control certain home appliances or loads from remote location is needed. The ICT gateway may also be connected to a market place to sell locally generated energy or get price signals indicating a low price for energy, which can be coupled with profiles for dedicated energy con-sumers. All these new services require a way for smart grid participants (e.g., distribution network provider, energy provider, marketplace, meter data management) to discover ICT gateways. Once discovered the secure connectivity to the user's home energy gateway provides the base for secure service provisioning and secure information exchange. Security comes in different flavors here, as at least authentication and authorization are required, to reliably control access to the ICT gateway and thus to secure energy appliances. Based on the required security level, integrity and privacy protection can leverage the initial security interaction. This supports also security requirements from different regulation bodies (examples are provided in [6] and [7]).

The remainder of this paper is organized as follows: Section II provides an overview of smart grid challenges with respect to security in general and describes in two subsections the end user and the utility perspective. Section III builds on that and describes security considerations to be obeyed when designing a secure access solution to home energy appliances. Section IV afterwards provides three different approaches to securely connect to the ICT gateway, whereby security is related to authentication and authorization of accessing peers. Section V concludes the paper.

## II. CHALLENGES IN THE SMART GRID

Current challenges for the power grid include the integration of fluctuating renewable energy sources, distributed power generation, short interval feedback on users on their energy usage, user indicated demand peaks, and the foreseeable need for the integration of electric vehicles, leading to an even higher energy demand of customers at peak times. A "smarter" grid can meet many of these challenges. With the move to a smart grid the importance of IT communication technologies in energy automation rises. With the availability of pervasive IT communication services, a variety of new use cases becomes possible enhancing the service to end customers and mitigates the impact of the challenges mentioned above.

Many use cases center around the Smart Home scenario. Smart Home in combination with the Smart Grid will allow people to understand how their household uses energy by smart metering, manage energy use better, e.g., through dynamic pricing or time of use pricing. Moreover, it enables them to sell locally generated energy on a market place, and reduce their carbon footprint. As the energy automation standard IEC 61850 (cf. [1]) is already used in the backend of energy automation networks, IEC 61850 is a good candidate to be used for communication between instances of the Smart Grid and the gateway of a Smart Home. The standard IEC 62351 (cf. [2] and [4]) defines security functions to protect IEC 61850 communication. The availability of an established security standard supports the approach to use IEC 61850 for communication between instances of the Smart Grid and the gateway of a Smart Home.

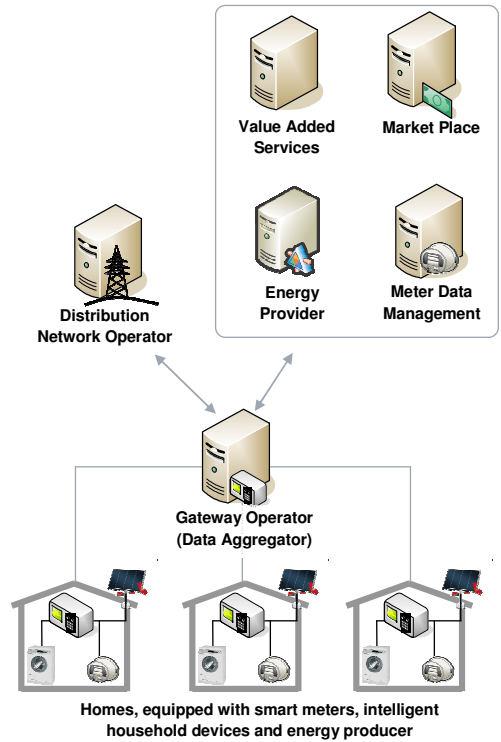

Figure 1: Connection of households to the smart grid

Figure 1 shows an example of a generic system architecture for a smart energy grid:

− Homes are equipped with smart meters, intelligent household devices, and energy producers.
− Home gateways control the communication between these entities within a home and with the smart grid and thus define a security perimeter. The ICT gateway hides the complexity of the in-house network from the smart grid. Moreover, by acting as a gateway, it also protects the user's privacy, in terms of installed devices, network topology, and the generated data, when communicating with other smart grid entities. The ICT gateway may also act as a proxy for the appliances of the home. An example is a connection to the energy market place typically requiring communication in both direc-tion, initiated by the household and the market place.

- A gateway operator is responsible for administration of the ICT gateways and provides connectivity for smart grid participants to the ICT gateways. As he controls the connectivity to the ICT gateway, a gateway operator may also provide centralized authorization functions for the ICT gateway. This also includes the user himself, when remotely controlling his household equipment.
- The distribution network operator is likely to communicate with the ICT gateways via the gateway operator, which hides the complexity of the ICT gateway management from the distribution network operator.
- The Meter Data Management processes the information received from the smart meters and provides the processed information to other market participants, e.g., for accounting.
- At the energy market, prosumers (resp. their ICT gateways) may buy and also sell energy. Hence the market offers a demand regulated price. An energy market alleviates the integration of distributed energy generators (e.g. solar cells).
- The smart grid information and communication infrastructure is also the enabler for other value added services. Remote serviceability of household equipment is just one example.

The functionality depicted can only be realized if two main issues are solved: one is the addressability through external entities (e.g. the Distribution Network Operator) - the ICT gateways must be reachable by external entities. The second, after being reachable, access to the ICT gateways must be restricted to a distinct group of people, like the market place sending offers, a smart grid operator sending commands to the gateway, or the user accessing the gateway remotely in order to meet security and privacy requirements.

*A. Consumer Perspective: Smart Home*

Having an information and communication infrastructure like this at hand, the following use cases can be supported in the user environment:

*1) Energy-aware home appliances*

Currently, the price of energy for private consumers is mostly constant. From the perspective of a utility it would be beneficial to have dynamic pricing to influence the energy usage of customers; therefore it is likely that there will be dynamic pricing in the near future. On the customer side, new intelligent, energy-aware home appliances can optimize the costs for energy usage by starting and stopping energy extensive tasks (e.g., cloth or dish washing, charging an electric car) at appropriate times (e.g., start when energy is cheap). This requires that the current price of energy is known and there is some way to determine the price of energy for the duration of an operation. One way to implement such a system is an energy market, where energy-aware home appliances buy a certain amount of energy before they start an operation. Especially charging a private electrical car during the night is an extremely flexible operation that requires much energy but has a large time window for execution, hence benefits from a good deal.

To integrate this functionality with the architecture presented above, the ICT gateway trades energy at the energy market. Accounting for any contract on the energy market includes the energy provider as well as the meter data management.

*2) Distributed power generation*

If energy is produced in a customer's home, e.g., by solar cells, the energy not needed at that point in time can be either stored or traded on an energy market. Especially if the energy market is on a large scale, selling the energy with dynamic pricing may be more attractive than fixed pricing. On the other hand, energy markets may be restricted to a smaller geographical area (e.g., city district) thus, energy can be locally distributed in the first place, supporting self-containment of that area.

As for the scenario above, the ICT gateway is connected with external smart grid entities to trade energy. Accounting for contracts includes the distribution network provider as well as the smart meter management.

*3) Energy Management and User Awareness*

An application with integrated user interface in the home is used for communication with the utility, e.g., to get a diagram of current energy usage, to get current energy pricing, to get the personal energy usage history, to get energy saving tips and the like. The user interface may also be used to receive energy outage forecasts, for troubleshooting, or to dynamically select a desired energy mix.

Even energy-aware home appliances may offer a user interface that states the current price for one operation execution. E.g. a coffee machine may state the price per coffee pot or the washing machine may suggest a different start time to save money because of a better tariff.

To implement this use case with the architecture presented above, the ICT gateway informs appliances about current energy prices, which it either gets at the energy market or directly from the energy provider (price signals as special incentive for a special behavior).

### B. Utility Perspective

Further use cases are focused on keeping the distribution network stable and keeping costs for utilities low (e.g., because it is not necessary to buy additional energy at short notice). As IEC 61850 is already widespread in use in the distribution network, it is a natural candidate for the following use cases:

#### 1) Reactive shutoff of home appliances

A utility has the ability to shut down certain home appliances in the household of users to react on certain situations in the network (e.g., if too many consumers are active). In doing so, the utility avoids buying energy on short notice, which is very expensive (prices may go up to 100 Euros per kWh). The ability and permission to issue switch-off commands to a user's home can be based on special contracts between user and utility operator. As the ICT gateway provides a security perimeter, it hides the internal energy consumer network from the utility.

To implement this use case with the architecture presented above, the utility must only have a certain amount of energy (kilowatt hours), which can be unloaded in households as well as the communication addresses of the associated ICT gateways. In the architecture above, ICT gateways may be addressed through the gateway operator that also ensures the connectivity of the ICT gateways. The utility sends a shutoff message via the gateway operator to a set of ICT gateways. Sending this shutoff message to many households must be finished in a short time to allow fast reactions. The shutoff message must be protected to avoid being misused by attackers. The ICT gateway takes the appropriate actions to meet the request of the utility, especially, it communicates with proper appliances to be shut off.

#### 2) Shutoff of power generator

The utility may not only tear down energy consumption, it may also instruct distributed power generators to not feed energy to the distribution network to fight situations when there is a low demand for energy. The signaling process is the same as in the previous use case. Again, the ability and permission to issue switch-off commands to power generators in the user's home is expected to be based on special contracts between user and utility operator.

#### 3) Demand Response

Another use case from a utility prospect is demand response: A utility can send price signals (either a rather high price if energy demand is too high or a low price if the energy demand is too low) without using the energy market to influence energy usage of intelligent home appliances. Price signals are especially interesting for energy intensive tasks that have a large time window for execution, e.g. charging of electric cars. Price signals can be sent for future time periods or as real time pricing information. The utility sends price signals via the gateway operator to the ICT gateways. The ICT gateway may distribute the pricing information in the home to the appropriate home appliances or act on the pricing information based on defined profiles.

### III. CONSIDERATIONS FOR ICT GATEWAY ACCESS

As described above, the secure discovery and access to a home energy gateway is a central functionality for a secure smart energy grid. Based on the described scenarios, the following entities may require secure access, via both local and remote communication, to an ICT gateway:

− Energy provider (including metering / billing)
− Distribution network operator
− Gateway operator
− Energy Market
− End user

Therefore an ICT gateway has to support remote access, allowing secure exchange of measurement, supervision and control data with the energy network. Secure access involves the following parts:

− Authentication: Mutual authentication between ICT gateway and accessing entity.

− Authorization: Determine access permissions. The decision whether a certain access is authorized, may be decided locally on the device or centrally by an authorization server. The access control en-

- Secure Communication: Confidentiality and Integrity of the communication is protected, e.g., by using IPsec or SSL/TLS.

forcement takes place at the gateway itself.

The following subsections discuss the different aspects for the realization of secure access to an ICT gateway and propose an appropriate solution.

### A. Platform Security

It is expected that the ICT gateway itself supports certain security features. These may comprise:

- Secure credential store for storing user or service relate credentials. These may be issued by the gateway service provider or even the energy provider himself. A flexible approach for providing a secure credential store is the use of pluggable security modules, e.g. in form of a smart card. Similar to a mobile phone SIM card, the user can then simply change his energy provider by replacing the energy provider's security module.
- Self-integrity check to ensure that the system configuration has not been altered in an unauthorized way
- Tamper protection to detect if changes to the hardware have been made.

The latter feature is especially important if the ICT gateway provides information for billing, is part of a smart meter device, or is mounted in a place, which are publically accessible.

### B. Network Connectivity

One requirement arising from these use cases is scalability. Security solutions for the Smart Grid must scale with millions of devices – Germany for example has more than 39 million households and each household is likely to have more than one device.

Users will connect their ICT gateways to already existing ICT networks, which are used for Internet access and services like Voice over IP (VoIP) or multimedia streaming. The user infrastructure typically involves a gateway connecting to the Internet (DSL box), which has a Network Address Translation (NAT) component as well as a firewall component integrated. Thus, accessibility of an ICT gateway connected to a user's home network is limited, as the ICT gateway is not addressable from outside. Hence, another requirement is NAT-traversal (the ability to address and reach a component behind a NAT device).

Multiple levels of hierarchy from a control station to a device in a household are a common solution to address scalability. This includes communication other than the point to point communication used today.

A solution that enables NAT-traversal is described below.

### C. ICT Gateway Discovery

As shown in Figure 1, in smart grid scenario's new roles and/or components may be introduced. One example is the ICT gateway operator. This gateway operator is in charge of concentrating the communication from the home energy gateways up to the control center as well as providing an easy way to the control center to reach a high number of energy gateways at once. Moreover, a gateway operator can support the discovery of ICT gateways and most likely will also offer additional services based on this discovery like remote access for user's or remote management of the ICT gateways, e.g., to provide enhanced functionality or updates for installed software.

### D. Permissions

The following permissions resp. kinds of access to the ICT gateway have to be distinguished:

o Change configuration

o Install applications

o Get status information

o Issue commands (e.g. shutoff command)

### E. Access Authentication

Access to the home energy gateway should be limited to a distinct group of entities, which may comprise the distribution network operator, the gateway operator, and the user itself. Moreover, communication with the gateway should be performed in a secure manner. Both can be achieved as described in the following:

- Gateways possess a certificate, which was either provided by the vendor of the gateway or the gateway operator.
- Roles or components accessing the gateway also possess a public resolvable certificate and corresponding private key.
- Mutual authentication between gateway and accessing component/role based on credentials is performed.
- Home energy gateways are not always reachable

from the Internet, i.e. they may not use a public IP address.

- The IP address may be resolved by the gateway operator directly or by a network operator (in case of DSL at the users home) if requested by other smart grid entities

In the following options for the access decision points are discussed.

*F. Local access control decision at the ICT gateway*

To ensure a secure communication of roles/persons allowed to access the home gateway, an access control list (containing credentials, potentially role information, associated rights, etc.) has to be managed at the gateway directly. This may be done by either the user or the gateway operator. As the user has the choice of a preferred energy provider, service aggregator or other smart grid roles, there will be individual communication connections per user. Thus user-specific configuration has to be done at each home energy gateway. Dedicated roles can initiate a secure connection to every ICT gateway based on a device certificate installed at the gateway.

The advantage of this approach is a direct accessibility for different roles or components to the ICT gateway. The disadvantage is that user specific connection credentials have to be administered at each ICT gateway separately. Note that the gateway discovery and address resolution is neglected here. Deployment in existing home environments may not lead to the establishment of a communication channel through the existence of NAT and Firewall functionality. Thus, the approach of direct access is not feasible.

*G. Centralized access control decision*

An alternative to the decentralized handling of access control is a central, network-based functionality of a security provider, who authenticates and authorizes potential roles and components, that wish to access an ICT gateway. Such an approach is already known from telecommunication environments like the Generic Bootstrapping Architecture (GBA) of 3GPP networks. A further example can be given through server components in voice over IP (VoIP) networks, e.g., in SIP (Session Initiation Protocol, RFC 3261). The following section investigates in centralized access control.

IV. SOLUTIONS FOR CENTRALIZED ACCESS CONTROL

The proposed central security function provides **A**u-thentication – **S**ession **I**nvocation – **A**uthorization. Hence the abbreviation ASIA is used further on for this component. ASIA would be typically combined with other functions needed, e.g., for the address resolution or ICT gateway discovery. It is therefore expected that the ICT gateways register with the ASIA component to publish their IP addresses. This registration connection is kept online to allow access to the ICT gateway from remote and may also be used to communicate consumption data or to connect to value added services. As it is expected that the ICT gateway is operated in a home environment behind the DSL router, this permanent connection also keeps the NAT bindings as well as the opened communication ports in the DSL router. To achieve connectivity with an ICT gateway using ASIA different modes are described in the following subsections.

*A. ASIA – Operation in Session Invocation Mode*

In this mode, ASIA initiates a connection establishment with requesting smart grid role/component.

Process:

- ASIA authenticates and authorizes the connecting smart grid role/component (energy provider, aggregator, meter data management, etc.)

- Accessing role/component provides a TAN (transaction number) for the ICT gateway to associate the ICT gateway in the direct connection later on. Alternatively, the TAN may be generated by ASIA and provided to the accessing role/component and the ICT gateway as part of a software token. This has the advantage that the software token may contain further information about security parameter of the connection to be established between the ICT gateway and the accessing role/component. Examples are the fingerprints of the used certificates for authentication at ASIA.

- ASIA sends the connection request from the role/component via the permanent connection to the ICT gateway including the connection parameter (address of role/component, TAN, software token, etc.) and optionally the requested command to be executed.

- The ICT gateway uses the received information to establish a direct connection to the requesting role/component. The distributed TAN can be used to associate the connection attempt with the initial request of the requesting role/component.

- After successful authentication as part of the direct connection, the information exchange can be started.

Technical approach:

- If web-based services are already used in the system the authorization process may be realized using SAML and XML security (cf. [3]). This approach supports certificate based authorization as well as pre-shared key based authorization.
- An alternative to SAML may be the application of Kerberos for a purely symmetric infrastructure without the need for certificates.
- Authentication on the direct connection may be performed on transport layer, e.g., by applying TLS.

The following figure shows the general protocol flow for a session invocation.

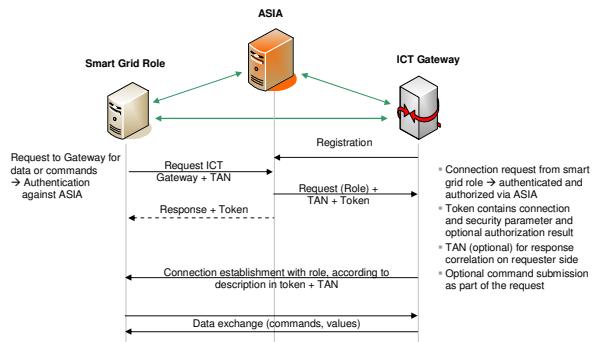

**Figure 2: ASIA – Session Invocation Mode**

The advantage of this approach is the flexibility and scalability in changing environments, especially in when the communication relations change. The approach supports a central authentication and authorization component, which keeps the administrative effort low per gateway. This may influence the used gateway hardware. The central component is not involved in the actual data exchange. The disadvantage is that there is no option for a direct communication establishment between the requesting role/component and the ICT gateway, which stems from the user environment, because of the missing TAN.

### B. ASIA – Operation in Redirect Mode

In this mode the ASIA server provides the address location information to the requestor after successful authentication. The requestor can then directly connect to the ICT gateway.

Process:

- The ASIA component authenticates the requesting smart grid role/component based upon certificates or pre-shared keys alternatively.
- According to the ASIA rule set, the address information will be provided to the requesting role/component. Here, it is also possible to include a software token as proof that the role/component has been authenticated and authorized at the ASIA server. This token may include further security parameter.

Technical approach:

Based on the described functionality, the ASIA server would resemble a Kerberos like KDC authenticating and authorizing the access to the ICT gateway. Note that this approach builds upon the general accessibility of the ICT gateway in terms of addressing and connection establishment, which may be limited due to the user's environment.

The following figure depicts the general call flow.

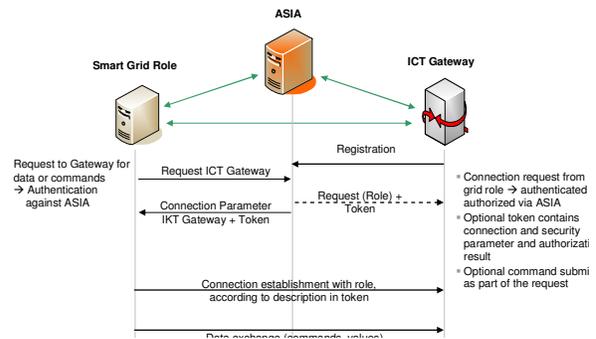

**Figure 3: ASIA – Redirect Mode**

The advantage of this approach is the flexibility to react on changing communication relations. Like the previous approach this solution allows for a central authentication and authorization component. The obvious disadvantage is the requirement to have direct accessibility to the ICT gateway from other components. In user environments with NAT and FW devices, this solution will fail, as the Firewall will block access to the ICT gateway. Moreover, if the user has a common DSL account the associated IP address is likely to change within 24 hours (at least in Germany).

### C. ASIA – Operation in Proxy Mode

The ASIA server may act as proxy for restricting access to the ICT gateway.

Process:

- The ASIA component authenticates the requesting role/component based upon certificates or pre-shared keys alternatively.
- Based on the permanent connection between the ICT gateway and the ASIA server, the connection attempt of the requesting role/component is forwarded to the ICT gateway.
- ASIA may provide information about the authentication and authorization state, in a software token.
- To achieve end-to-end security also in the case of a communication proxy, application layer security mechanisms may be used. An example is the application of XML signatures to achieve integrity protection.

Technical approach:

- ASIA simply connects both links and operates as a data proxy for the duration of the connection.

This approach may be realized by using the previously described approach, with the exception, that ASIA does deliver the own address instead of providing the ICT gateway address. It may be realized with the first approach letting the ASIA server connect to the requesting role/component. Thus, this approach is rather a configuration of the first two approaches with central access control.

The following figure shows the abstract call flow or the connection establishment.

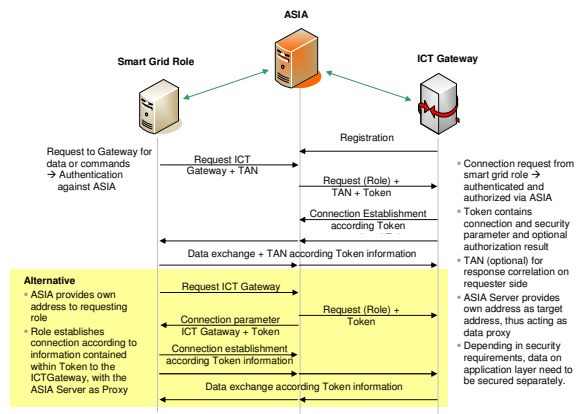

Figure 4: ASIA – Proxy Mode

The advantage of this approach is the flexibility to react on changing communication relations. A disadvantage is the missing end to end connection establishment. Moreover, as all traffic is processed though the ASIA server, dedicated hardware requirements may arise.

Generally, the ASIA functionality can also be used to deliver configuration information, which communication connections are allowed or normal and which not. This configuration information can also be used to enrich the effectiveness of IDS/IPS tools, by making them aware of the network configuration.

V. CONCLUSION

This paper provides an overview of smart grid environment and challenges. Many future use cases in the smart grid are centered on the smart home. An intelligent ICT gateway is the key to provide the intended services. Nevertheless, the base for the service provisioning is connectivity. This paper described approaches to protect access and data exchange, preventing manipulation of ICT gateway operation. It presented an ASIA (Authentication, Session Invocation, Authorization) component to be used in the smart grid environments to protect access to energy appliances at a user's home from remote and to cope with basic problems like ICT gateway discovery and addressing. Three different modes of operation allow the ASIA component to be adapted to a very large range of smart grid communication network settings. The concept of ASIA was also introduced into the E-DeMa project (cf. [5]), which is one of six projects funded by the German Federal Ministry of Economics and Technologies (BMWi) in the national joint research initiative E-Energy.